\documentclass[12pt]{article}

\usepackage{graphicx,amssymb,amsmath,dsfont,txfonts} 
\usepackage{bbm} 
\usepackage{mathtools} 

\usepackage[a4paper]{geometry} 
\def\dfrac#1#2{\frac{\displaystyle\strut #1}{\displaystyle\strut #2}}
\DeclareMathOperator{\tr}{tr}
\def\bra#1{\mathinner{\langle{#1}|}}
\def\ket#1{\mathinner{|{#1}\rangle}}
\def\braket#1{\mathinner{\langle{#1}\rangle}}

\begin{document}

\title{\LARGE\bf 
Indefinite causal order for quantum phase estimation with Pauli noise}

\author{Fran\c{c}ois {\sc Chapeau-Blondeau}, \\
    Laboratoire Angevin de Recherche en Ing\'enierie des Syst\`emes (LARIS), \\
    Universit\'e d'Angers,
    62 avenue Notre Dame du Lac, 49000 Angers, France.
}


\maketitle

\parindent=8mm \parskip=0ex

\begin{abstract}
This letter further explores the recent scheme of switched quantum channels with
indefinite causal order applied to the reference metrological task of quantum phase estimation 
in the presence of noise. We especially extend the explorations, previously reported with 
depolarizing noise and thermal noise, to the class of Pauli noises, important to the qubit and 
not previously addressed. Nonstandard capabilities, not accessible with standard quantum phase 
estimation, are exhibited and analyzed, with significant properties that are specific to the 
Pauli noises, while other properties are found in common with the depolarizing noise or the 
thermal noise. The results show that the presence and the type of quantum noise are both crucial 
to the determination of the nonstandard capabilities from the switched channel with indefinite 
causal order, with a constructive action of noise reminiscent of stochastic resonance phenomena.
The study contributes to a more comprehensive and systematic characterization of the roles and 
specificities of quantum noise in the operation of the novel devices of switched quantum channels 
with indefinite causal order.
\end{abstract}

\maketitle

\section{Introduction}

{\let\thefootnote\relax\footnote{{Preprint of a paper published by {\em Fluctuation and Noise Letters},
vol.~22, 2350036, pp.~1--9 (2023). \\
https://doi.org/10.1142/S0219477523500360 }}}
A quantum signal placed in a coherent superposition can be used to switch between two possible 
causal orders for cascading two quantum processes or channels. This realizes a switched composite 
channel presenting simultaneously the two possible causal orders of the two cascaded quantum 
processes, or exhibiting indefinite causal order. The scheme is illustrated in Fig.~\ref{figSwiP1}. 
Such quantum superposition with indefinite causal order of quantum channels, recently introduced in 
\cite{Oreshkov12,Chiribella13} and experimentally explored in \cite{Procopio15,Goswami18,Guo20},
in addition to the fundamental issues it raises concerning the nature of physical causality, is 
progressively recognized for its potential to quantum information processing. 
The scheme has first been shown exploitable to enhance the information capacity of noisy quantum 
channels \cite{Ebler18,Procopio19}, and more recently for enhanced performance in noisy quantum 
metrology \cite{Chapeau21,Chapeau22}. For further exploration in quantum metrology, here we 
investigate the reference task of quantum phase estimation in the presence of noise, when the quantum 
process upon which the phase is to be estimated, is duplicated and inserted in a switched quantum 
channel with indefinite causal order under the control of a qubit signal. Several nonstandard 
properties relevant to estimation arise, that are not accessible with standard quantum phase
estimation. Such properties were first reported in \cite{Chapeau21,Chapeau22} with some specific 
qubit noise models, namely depolarizing noise in \cite{Chapeau21} and thermal noise in 
\cite{Chapeau22}. Here, we extend the exploration and analysis to other noise models important to 
the qubit, under the form of Pauli noises, not explored in \cite{Chapeau21,Chapeau22}.
The present report discloses new results not contained in \cite{Chapeau21,Chapeau22}, and it 
shows the importance of the noise, its level and properties, for obtaining coherent superposition 
of causal orders and their non-standard capabilities relevant to phase estimation,
with significant properties specific to the Pauli noises.

We consider the generic metrological scenario \cite{Chapeau16} where a probe qubit with density 
operator $\rho =\bigl( \mathrm{I}_2 + \vec{r}\cdot \vec{\sigma} \bigr)/2$ in standard Bloch 
representation \cite{Nielsen00}, is transformed by the unitary operator 
$\mathsf{U}_\xi=\exp\bigl(-i \xi \vec{n} \cdot \vec{\sigma} /2 \bigr)$,
and subsequently affected by a quantum noise $\mathcal{N}(\cdot)$, so as to implement the
overall quantum process or channel
$\rho \mapsto \mathcal{E}_\xi(\rho)=\mathcal{N}\bigl( \mathsf{U}_\xi \rho \mathsf{U}_\xi^\dagger \bigr)$.
The noisy probe qubit in state $\mathcal{E}_\xi(\rho)$ is to be used to estimate the phase $\xi$ 
of the unitary $\mathsf{U}_\xi$.

Standard phase estimation \cite{Chapeau16} from the noisy probe state
$\mathcal{E}_\xi(\rho)$, for good efficiency, relies on specific properties.
(i) The input probe $\rho$ must be a pure state with a Bloch vector of unit norm
$\| \vec{r}\, \|=1$, and with in $\mathbbm{R}^3$ an orientation $\vec{r}$ orthogonal
to the unit vector $\vec{n}$ characterizing the unitary $\mathsf{U}_\xi$.
(ii) The output probe qubit must be measured in directions matched to
both the orientations of the unitary $\vec{n}$ and of the output density operator
$\mathcal{E}_\xi(\rho)$ (which is typically dependent on the unknown phase $\xi$).
(iii) The estimation efficiency usually degrades monotonically as the level of the quantum
noise $\mathcal{N}(\cdot)$ increases.
By contrast, nonstandard estimation when the process $\mathcal{E}_\xi(\cdot)$ is inserted in a 
switched quantum channel with indefinite causal order, is not limited in the same way by
these properties, as we are going to see.

\section{Switched channel with indefinite causal order}

\begin{figure}[htb]
\centerline{\includegraphics[width=70mm]{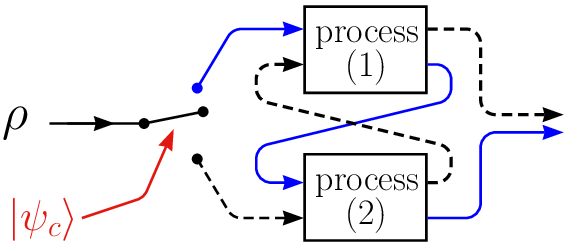}} 
\caption[what appears in lof LL p177]
{Switched quantum channel where, for transmitting the probe qubit $\rho$, two quantum processes 
(1) and (2) can be cascaded either in the causal order (1)-(2) (solid path) or in the causal order 
(2)-(1) (dashed path), according to the state, respectively $\ket{\psi_c}=\ket{0_c}$ or
$\ket{\psi_c}=\ket{1_c}$, of a control qubit. When the control qubit is placed in a superposed 
state $\ket{\psi_c}$ as in Eq.~(\ref{psic_super}), a coherent superposition of the two causal 
orders (1)-(2) and (2)-(1) is realized and experienced by the probe qubit $\rho$ in its
transmission. In the present study, each process (1) and (2) is realized by a copy of the process 
$\mathcal{E}_\xi(\cdot)=\mathcal{N}\bigl( \mathsf{U}_\xi \cdot \mathsf{U}_\xi^\dagger \bigr)$.
}
\label{figSwiP1}
\end{figure}

In the switched quantum channel of Fig.~\ref{figSwiP1}, we consider the situation where each 
quantum process (1) and (2) is the process 
$\rho \mapsto \mathcal{E}_\xi(\rho)=\mathcal{N}\bigl( \mathsf{U}_\xi \rho \mathsf{U}_\xi^\dagger \bigr)$
realized by the same unitary $\mathsf{U}_\xi$ affected by an independent realization of the noise 
$\mathcal{N}(\cdot)$. One copy of $\mathcal{E}_\xi(\cdot)$ for process (1) and one copy of 
$\mathcal{E}_\xi(\cdot)$ for process (2) are assembled to form the switched quantum channel of 
Fig.~\ref{figSwiP1}, under the control of a qubit signal in a state $\ket{\psi_c}$.

With a control signal $\ket{\psi_c}$ in the basis state $\ket{\psi_c}=\ket{0_c}$ or 
$\ket{\psi_c}=\ket{1_c}$, a single causal order is realized in Fig.~\ref{figSwiP1}, with the 
switched channel implementing on the input probe $\rho$ the standard cascade 
$\rho \mapsto \mathcal{E}_\xi \circ \mathcal{E}_\xi(\rho)$. Estimation of the phase $\xi$ from the state 
$\mathcal{E}_\xi \circ \mathcal{E}_\xi(\rho)$, although the phase experienced by the probe is amplified 
to $2\xi$ and altered twice by the noise process $\mathcal{N}(\cdot)$, is in essence no different
from standard estimation with the same limiting properties as mentioned in the Introduction.

Novel nonstandard properties arise when the control qubit is placed in a superposed state
\begin{equation}
\ket{\psi_c}=\sqrt{p_c}\ket{0_c}+\sqrt{1-p_c}\ket{1_c} \;,
\label{psic_super}
\end{equation}
with a probability $p_c \not = 0, 1$. This realizes in Fig.~\ref{figSwiP1} a coherent superposition 
of two possible causal orders of the cascaded processes, or an indefinite causal order, experienced 
by the probe qubit across the noisy processes $\mathcal{E}_\xi(\cdot)$.

With the control density operator $\rho_c=\ket{\psi_c}\bra{\psi_c}$, Ref.~\cite{Chapeau21}
works out a general characterization of the action of the switched quantum channel of 
Fig.~\ref{figSwiP1}, which is found to realize the bipartite transformation of the probe-control 
qubit pair described by the superoperator
\begin{equation}
\mathcal{S}(\rho \otimes \rho_c) = \mathcal{S}_{00}(\rho) \otimes 
\Bigl[ p_c \ket{0_c}\!\bra{0_c} + (1-p_c) \ket{1_c}\!\bra{1_c} \Bigr] + \mathcal{S}_{01}(\rho) 
\otimes \sqrt{(1-p_c)p_c} \, \Bigl( \ket{0_c}\!\bra{1_c} + \ket{1_c}\!\bra{0_c} \Bigr) \;.
\label{Sgenqb}
\end{equation}

In Eq.~(\ref{Sgenqb}), the superoperator 
$\mathcal{S}_{00}(\rho) \equiv \mathcal{E}_\xi \circ \mathcal{E}_\xi(\rho)$ represents the density 
operator produced by the standard cascade alone. Meanwhile, $\mathcal{S}_{01}(\rho)$ is a superoperator
conveying the effect of indefinite causal order in the switched channel of Fig.~\ref{figSwiP1}, and 
acting only at $p_c \not = 0, 1$, as visible from Eq.~(\ref{Sgenqb}). It is defined \cite{Chapeau21} 
as $\mathcal{S}_{01}(\rho) = \sum_{j,k} \mathsf{K}_j \mathsf{K}_k \rho 
\mathsf{K}_j^\dagger \mathsf{K}_k^\dagger$ where the $\mathsf{K}_j$'s are the Kraus operators
of the process 
$\mathcal{E}_\xi(\cdot)=\mathcal{N}\bigl( \mathsf{U}_\xi \cdot \mathsf{U}_\xi^\dagger \bigr)$
determined by the quantum noise $\mathcal{N}(\cdot)$ and the unitary $\mathsf{U}_\xi$.

To apply the joint state $\mathcal{S}(\rho \otimes \rho_c)$ of Eq.~(\ref{Sgenqb}) in a task 
of quantum phase estimation in the presence of noise, for the noise process
$\mathcal{N}(\cdot)$, Ref.~\cite{Chapeau21} considered the highly symmetric, isotropic, 
depolarizing noise model \cite{Nielsen00}, while Ref.~\cite{Chapeau22} studied a less symmetric qubit 
thermal noise model \cite{Chapeau22b}, and Refs.~\cite{Chapeau21,Chapeau22} both reported nonstandard 
properties for phase estimation from the switched quantum channel. Here we extend the exploration to 
other noise models, not explored in \cite{Chapeau21,Chapeau22}, and important to the qubit, consisting 
in the class of Pauli noises, that for instance includes the bit-flip and the phase-flip noises.

A property with a strong nonstandard character afforded by the switched channel, and on which we 
concentrate in the sequel, is that phase estimation
can be performed by measuring the control qubit alone. The control qubit in the switched channel 
of Fig.~\ref{figSwiP1} never directly interacts with the unitary $\mathsf{U}_\xi$ under estimation. 
It is the probe qubit alone that interacts with $\mathsf{U}_\xi$. Nevertheless, the type of 
coupling taking place in the switched channel transfers information about the phase $\xi$ to the
control qubit. This can be seen from the bipartite state $\mathcal{S}(\rho \otimes \rho_c)$ of 
Eq.~(\ref{Sgenqb}), via partial tracing over the probe, when the control qubit alone is to be 
used for estimation, to give 
$\rho^{\rm con}=\tr_{\rm probe}\bigl[ \mathcal{S}(\rho \otimes \rho_c) \bigr]$.
From Eq.~(\ref{Sgenqb}), this leads to the reduced state of the control qubit
\begin{equation}
\rho^{\rm con} = p_c \ket{0_c}\!\bra{0_c} + (1-p_c) \ket{1_c}\!\bra{1_c} 
+ Q_c(\xi) \sqrt{(1-p_c)p_c} \Bigl( \ket{0_c}\!\bra{1_c} + \ket{1_c}\!\bra{0_c} \Bigr) \;,
\label{Sgenqb_tp1}
\end{equation}
with $Q_c(\xi)=\tr[\mathcal{S}_{01}(\rho)]$.
By contrast, if the probe qubit alone were to be used for estimation, partial tracing over the 
control in Eq.~(\ref{Sgenqb}) would give for the probe the reduced state 
$\tr_{\rm control}\bigl[ \mathcal{S}(\rho \otimes \rho_c) \bigr] =\mathcal{S}_{00}(\rho)$, that
is nothing else than a probing of the standard cascade 
$\mathcal{S}_{00}(\rho) \equiv \mathcal{E}_\xi \circ \mathcal{E}_\xi(\rho)$.
When the control qubit is not exploited at the output, the switched channel reduces to the 
standard cascade $\mathcal{E}_\xi \circ \mathcal{E}_\xi(\rho)$.

Proceeding with the possibility of exploiting the control qubit alone for phase estimation,
to explicitly evaluate the factor $Q_c(\xi)=\tr[\mathcal{S}_{01}(\rho)]$ for the control signal
of Eq.~(\ref{Sgenqb_tp1}), we specify, as announced, to the Pauli noises for $\mathcal{N}(\cdot)$,
implementing on a generic state $\rho$ the nonunitary transformation
\begin{equation}
\rho \longmapsto \mathcal{N}(\rho)= (1-p)\rho + p \sigma_\ell \rho \sigma_\ell^\dagger \;.
\label{PauliN}
\end{equation} 
The subscript $\ell =x, y, z$ identifies one of the three Pauli operators $\sigma_\ell$. In this 
way, the quantum noise in Eq.~(\ref{PauliN}) applies with a probability $p$ the Pauli operator 
$\sigma_\ell$ to alter the qubit state. When $\ell =x$, the noise applies $\sigma_x$ and is a 
bit-flip noise; when $\ell =z$, it is a phase-flip noise; and when $\ell =y$, a bit-phase-flip 
noise \cite{Nielsen00}. This determines the two Kraus operators 
$\mathsf{K}_1=\sqrt{1-p}\mathsf{U}_\xi$ and $\mathsf{K}_2=\sqrt{p} \sigma_\ell \mathsf{U}_\xi$ for 
$\mathcal{E}_\xi(\cdot)$ and then yields
\begin{equation}
Q_c(\xi)= 1-2(1-n_\ell^2)(1-p)p\bigl[1-\cos(\xi)\bigr] \;,
\label{Qxi1}
\end{equation}
where $n_\ell$ is the component of the unitary axis $\vec{n}$ in the direction $\ell$ of
$\mathbbm{R}^3$ selected by the Pauli noise. Via $Q_c(\xi)$ we obtain an explicit dependence on the 
phase $\xi$ of the reduced state $\rho^{\rm con}$ of the control qubit, which can then be measured 
for estimating $\xi$.

To evaluate the performance, a meaningful and standard criterion is the quantum Fisher information 
$F_q^{\rm con}(\xi)$ contained in the state $\rho^{\rm con}$ about the phase $\xi$, which
quantifies the overall best estimation efficiency \cite{Demkowicz14,Chapeau16}. It is obtained as
\begin{equation}
F_q^{\rm con}(\xi) = 4(1-p_c)p_c \dfrac{\,\bigl[\partial_\xi Q_c(\xi) \bigr]^2}{1-Q_c^2(\xi)} \;.
\label{Fc3_b}
\end{equation}
It then follows that $F_q^{\rm con}(\xi)$ in Eq.~(\ref{Fc3_b}) is maximized at $p_c=1/2$, when the 
two causal orders superposed in Fig.~\ref{figSwiP1} have equal weight, to give the maximum
\begin{equation}
F_q^{\rm con}(\xi) = \frac{\bigl[2(1-n_\ell^2)(1-p)p\sin(\xi)\bigr]^2}
{1-\bigl\{1-2(1-n_\ell^2)(1-p)p[1-\cos(\xi)]\bigr\}^2} \;.
\label{Fc3_c}
\end{equation}
Another remarkable property, not always granted with standard phase estimation \cite{Chapeau16}, 
is that there exists a fixed, phase-independent, optimal measurement of the control qubit, by 
means of a projective von Neumann measurement in the Hadamard orthonormal basis 
$\bigl\{\ket{+}, \ket{-}\bigr\}$. The two possible measurement outcomes have probabilities
\begin{equation}
P^{\rm con}_\pm =\braket{\pm | \rho^{\rm con} | \pm}=
\frac{1}{2} \pm \sqrt{(1-p_c)p_c} \, Q_c(\xi) \;,
\label{Pc+a}
\end{equation}
associated with the classical Fisher information
\begin{equation}
F_c^{\rm con}(\xi)= \dfrac{(\partial_\xi P^{\rm con}_+)^2}{(1-P^{\rm con}_+)P^{\rm con}_+} \;,
\label{Fc2}
\end{equation}
which at the optimum $p_c=1/2$ gets maximized at the level of $F_q^{\rm con}(\xi)$ of Eq.~(\ref{Fc3_c}), 
establishing in this way the optimality of the measurement.

\begin{figure}[htb]
\centerline{\includegraphics[width=84mm]{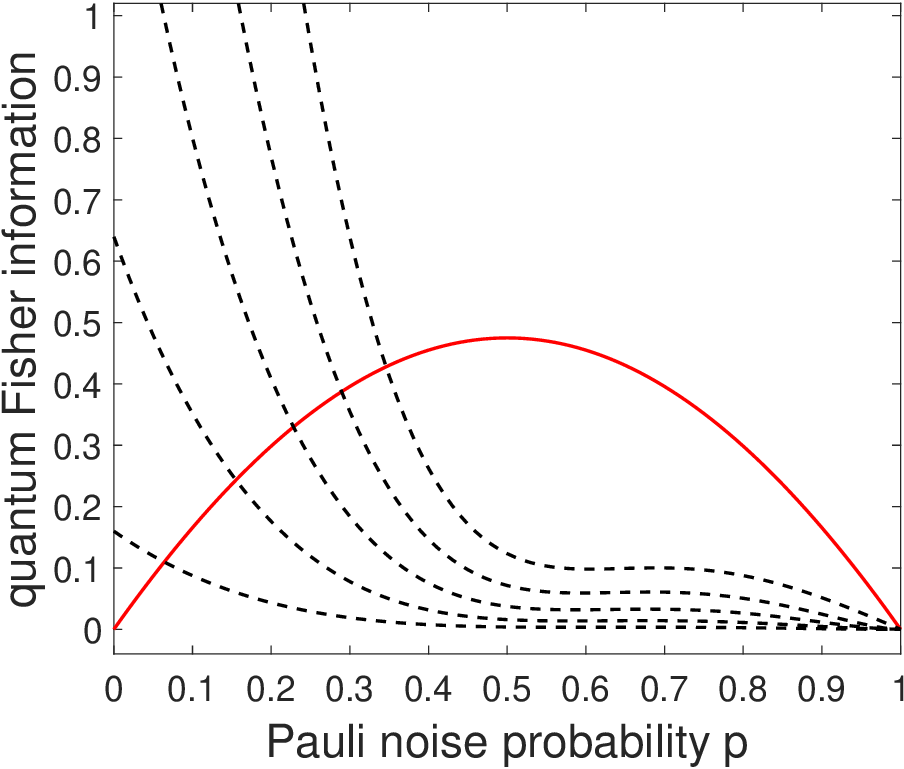}}
\caption[what appears in lof LL p177]
{As a function of the probability $p$ quantifying the level of the Pauli noise in
Eq.~(\ref{PauliN}), the quantum Fisher information $F_q^{\rm con}(\xi)$ from Eq.~(\ref{Fc3_c}) 
(solid line) accessible by measuring the control qubit of the switched channel of
Fig.~\ref{figSwiP1}. 
The $5$ curves in dashed lines show the quantum Fisher information $F_q^{\rm cas}(\xi \,; r)$ 
for the standard cascade $\mathcal{E}_\xi \circ \mathcal{E}_\xi(\rho)$ with an input probe signal
with (in decreasing order from the top to the bottom curve) $\| \vec{r}\, \|=r=1, 0.8, 0.6, 0.4$ 
and $r=0.2$.
}
\label{fig_Fq1}
\end{figure}

For illustration of typical nonstandard properties that can be obtained for estimation from
the switched channel of Fig.~\ref{figSwiP1}, we consider a unitary $\mathsf{U}_\xi$ with axis 
$\vec{n}=[0, 1, 0]^\top = \vec{e}_y$ and phase $\xi =\pi /5$.
The input probe is chosen with a Bloch vector $\vec{r}$ in the plane $(Ox, Oz)$ orthogonal to the 
unitary axis $\vec{n}$ in $\mathbbm{R}^3$, which stands as a necessary condition of
optimality for the phase estimation \cite{Chapeau16}. We take (illustratively) 
$\vec{r}=[0, 0, r]^\top = r\vec{e}_z$. The Pauli quantum noise $\mathcal{N}(\cdot)$ in
Eq.~(\ref{PauliN}) is taken as a bit-flip noise $\ell=x$. A phase-flip noise with $\ell=z$ 
would lead in Eq.~(\ref{Fc3_c}) to a same efficiency $F_q^{\rm con}(\xi)$ for estimating the phase 
$\xi$, with a same $n_\ell =0$. 
Figure~\ref{fig_Fq1} shows the quantum Fisher information $F_q^{\rm con}(\xi)$ from Eq.~(\ref{Fc3_c}) 
accessible by measuring the control qubit of the switched channel of Fig.~\ref{figSwiP1}, 
along with a comparison with the quantum Fisher information $F_q^{\rm cas}(\xi \,; r)$ for the standard 
cascade $\mathcal{E}_\xi \circ \mathcal{E}_\xi(\rho)$ as deduced from \cite{Chapeau16}. 

\smallbreak

\section{Discussion}

We emphasize that for the standard cascade compared in Fig.~\ref{fig_Fq1}, there may not be
\cite{Chapeau16} a phase-independent measurement to reach the quantum Fisher information 
$F_q^{\rm cas}(\xi \,; r)$, while this is always achieved for $F_q^{\rm con}(\xi)$ by measuring the 
control qubit in the fixed basis $\bigl\{\ket{+}, \ket{-}\bigr\}$.
A significant nonstandard property for estimation illustrated in Fig.~\ref{fig_Fq1}, is that 
the efficiency $F_q^{\rm con}(\xi)$ in Eq.~(\ref{Fc3_c}) of the control qubit, is unaffected by 
the input probe $\vec{r}$, and the same efficiency $F_q^{\rm con}(\xi)$ is preserved as the input 
probe degrades or depolarizes as $\| \vec{r}\, \| \rightarrow 0$ or tends to align with the unitary 
axis $\vec{n}$. In such conditions, estimation from the standard cascade 
$\mathcal{E}_\xi \circ \mathcal{E}_\xi(\rho)$ has an efficiency $F_q^{\rm cas}(\xi \,; r)$ 
following from \cite{Chapeau16} that diminishes and goes to zero, as visible in Fig.~\ref{fig_Fq1}.
Also in Fig.~\ref{fig_Fq1}, at small noise level $p$ and with a pure 
input probe $\| \vec{r}\, \| \rightarrow 1$, the standard cascade displays an efficiency 
$F_q^{\rm cas}(\xi \,; r)$ superior to that of the control qubit $F_q^{\rm con}(\xi)$, yet with
$F_q^{\rm cas}(\xi \,; r)$ that monotonically degrades as the noise level $p$ increases.
By contrast, the efficiency $F_q^{\rm con}(\xi)$ of the control qubit experiences a nonmonotonic
evolution as the noise level $p$ increases, culminating at a maximum for an intermediate noise
level. The quantum noise is not univocally detrimental, as it serves, in the switched channel, to 
couple the control qubit to the phase $\xi$. With no noise at $p=0$, the two superposed processes 
in Fig.~\ref{figSwiP1} are two identical indistinguishable unitaries $\mathsf{U}_\xi$, and the 
switched channel with $\mathcal{S}_{01}(\rho) =\mathcal{S}_{00}(\rho)$ reduces to the standard 
cascade $\mathcal{E}_\xi \circ \mathcal{E}_\xi(\rho)$, governed by 
$\mathcal{S}(\rho \otimes \rho_c) = \mathcal{S}_{00}(\rho) \otimes \rho_c$ in Eq.~(\ref{Sgenqb}),
and no coupling of the control qubit to the phase $\xi$ occurs. The noise is necessary 
to implement a superposition of two distinct causal orders in Fig.~\ref{figSwiP1}, inducing a
coupling expressed by Eq.~(\ref{Sgenqb}) of the control qubit to the unitary $\mathsf{U}_\xi$
and its phase $\xi$. As a result, at large noise level $p$ in Fig.~\ref{fig_Fq1}, the
control qubit of the switched channel has an efficiency $F_q^{\rm con}(\xi)$ always larger
than $F_q^{\rm cas}(\xi \,; r)$ of the standard cascade.
A maximum efficiency $F_q^{\rm con}(\xi)$ of the control qubit is obtained at the intermediate 
noise level $p=1/2$ in Fig.~\ref{fig_Fq1}. At the other extreme, when the noise level 
$p\rightarrow 1$, the two superposed processes in Fig.~\ref{figSwiP1} again tend to two identical 
indistinguishable unitaries $\sigma_\ell \mathsf{U}_\xi$, and no coherent superposition of two 
distinct causal orders occurs; no coupling of the control qubit with the unitary
$\mathsf{U}_\xi$ takes place in Eqs.~(\ref{Sgenqb_tp1}) and (\ref{Qxi1}), and a 
vanishing efficiency $F_q^{\rm con}(\xi)$ follows in Eq.~(\ref{Fc3_c}).
This type of constructive nonmonotonic action from the noise is reminiscent of a phenomenon of 
stochastic resonance or noise-enhanced processing, which has been reported and analyzed under 
many forms in the classical domain \cite{Chapeau96d,Kish01,Rousseau02,McDonnell08}, and more
recently in the quantum domain \cite{Bowen06,Wilde09k,Chapeau15c}. Here, in the quantum domain, 
in a comparable way, the switched channel benefits from a constructive implication of quantum 
noise or decoherence, which becomes favorable to the estimation efficiency, as also observed in 
other situations in \cite{Chapeau15c,Gillard19}.

We observe that the limiting properties (i)--(iii) of standard phase estimation mentioned in 
the Introduction, do not affect in the same way estimation from the switched channel, that often
offers distinctive complementary capabilities. The results here extend to Pauli noises the 
demonstration of nonstandard capabilities useful to phase estimation by exploiting the control 
qubit of the switched channel, with also specific properties not present with the other noise 
models previously examined in this context \cite{Chapeau21,Chapeau22}.
By comparison, the isotropic depolarizing noise in \cite{Chapeau21}, with the highest symmetry, 
leads to an efficiency $F_q^{\rm con}(\xi)$ of the control qubit for phase estimation, that is 
independent of both the input probe $\vec{r}$ and of the axis $\vec{n}$ of the unitary.
With the thermal noise in \cite{Chapeau22}, the efficiency $F_q^{\rm con}(\xi)$ does depend on 
the input probe $\vec{r}$, but is independent of the unitary axis $\vec{n}$. Conversely
with Pauli noises here, we find that the efficiency $F_q^{\rm con}(\xi)$ is independent of the 
input probe $\vec{r}$, but it depends on the unitary axis $\vec{n}$ via its component
$n_\ell$ in the direction $\ell$ selected by the Pauli noise, as shown by Eq.~(\ref{Fc3_c}). 
In particular, configurations with $n_\ell = \pm 1$ identify a unitary axis $\vec{n}$ aligned 
with the direction $\ell$ set by the Pauli noise, and this cancels the dependence with the phase 
$\xi$ of the coupling $Q_c(\xi)$ in Eq.~(\ref{Qxi1}) and thus sets to zero the Fisher information 
$F_q^{\rm con}(\xi)$ in Eq.~(\ref{Fc3_c}). In such configurations with $n_\ell = \pm 1$, the 
unitary operator reduces to $\mathsf{U}_\xi=\exp\bigl(\pm i \xi \sigma_\ell /2 \bigr)$ and comes 
to commute with the Kraus operator $\sqrt{p}\sigma_\ell$ of the Pauli noise. The Kraus operators
$\bigl\{\mathsf{K}_1=\sqrt{1-p}\mathsf{U}_\xi, \mathsf{K}_2=\sqrt{p}\sigma_\ell \mathsf{U}_\xi \bigr\}$
defining each superposed process (1) and (2) in Fig.~\ref{figSwiP1} then all commute. With 
commuting Kraus operators, the superoperator 
$\mathcal{S}_{01}(\rho) = \sum_{j,k} \mathsf{K}_j \mathsf{K}_k \rho \mathsf{K}_j^\dagger 
\mathsf{K}_k^\dagger$ intervening in Eq.~(\ref{Sgenqb}) to determine the superposition
of causal orders in the switched channel, comes to coincide with $\mathcal{S}_{00}(\rho)$.
As we have seen, when $\mathcal{S}_{01}(\rho)=\mathcal{S}_{00}(\rho)$ the switched channel
is no different from the standard cascade 
$\mathcal{S}_{00}(\rho) \equiv \mathcal{E}_\xi \circ \mathcal{E}_\xi(\rho)$, no superposition
of two distinct causal orders takes place in Fig.~\ref{figSwiP1}, and no coupling
of the control qubit to the unitary $\mathsf{U}_\xi$ occurs.
The importance of non-commuting Kraus operators for obtaining coherent superposition
of causal orders was recognized early in \cite{Procopio15,Ebler18}.
By contrast, with the depolarizing noise of \cite{Chapeau21} and the thermal noise of 
\cite{Chapeau22}, no configurations exist where the Kraus operators all commute, which makes a 
notable difference for the switched channel.
These behaviors reinforce the notion that the noise, via its Kraus operators, is essential to 
determine the conditions of coupling of the control qubit to the unitary $\mathsf{U}_\xi$.
Consistently, as the noise properties significantly vary, when moving from the depolarizing noise 
in \cite{Chapeau21} and thermal noise in \cite{Chapeau22} to the 
Pauli noises here, significant properties of the switched channel are modified, as noted above.
Some properties are observed in common with the noise models of \cite{Chapeau21,Chapeau22} 
and the Pauli noises here, especially that the performance of the control qubit for phase 
estimation depends nonmonotonically on the level of noise and that a nonzero optimal level of 
noise maximizes the efficiency. By contrast, other properties like the dependence or independence 
of the efficiency $F_q^{\rm con}(\xi)$ on the unitary axis $\vec{n}$ and input probe $\vec{r}$ are 
specific and change according to the noise model. In this respect, the present study contributes 
to a more comprehensive and systematic characterization of the roles and specificities of quantum 
noise in the operation of the novel devices of switched quantum channels with indefinite causal 
order.

In the switched channel, from a control qubit that does not directly interact with the unitary 
$\mathsf{U}_\xi$, estimation of the phase $\xi$ can be achieved, even with an input probe fully 
depolarized when $\| \vec{r}\, \|=0$ or aligned with the unitary axis when 
$\vec{r} \varparallel \vec{n}$, while this is generally not possible with standard estimation as 
discussed in the Introduction. Ancilla qubits that do not actively interact with the process under 
estimation can also be exploited by standard estimation, but they always need to be jointly 
measured with the active probing qubits to be of some utility for estimation \cite{Demkowicz14}, 
while here the control qubit can be measured alone for estimation.

Other properties of switched quantum channels with indefinite causal order can be investigated.
For instance here the contribution of jointly measuring the probe-control qubit pair at the 
output can be examined based on Eq.~(\ref{Sgenqb}), and it will usually combine the more 
standard properties from the probe qubit with the nonstandard properties contributed by the 
control qubit. Beyond, many aspects of quantum superpositions with indefinite causal order 
remain to be explored, for quantum information processing, quantum technologies and quantum 
engineering \cite{Nie22}.


\end{document}